\documentclass{aastex62}
\usepackage{amsmath}
\begin{document}

\title{Energy Balance in Avalanche Models for Solar Flares}

\author{Nastaran Farhang}
\affil{Department of Physics, Faculty of Science, University of Zanjan \\
 P.O. Box 45195-313, Zanjan, Iran}
\email{[farhangnastaran,safari]@znu.ac.ir}
\author{Michael S. Wheatland}
\affiliation{Sydney Institute for Astronomy, School of Physics, The University of Sydney \\
 NSW 2006, Australia}

\author{Hossein Safari}
\affiliation{Department of Physics, Faculty of Science, University of Zanjan \\
 P.O. Box 45195-313, Zanjan, Iran}

\begin{abstract}
The distributions of solar flare energies and waiting times have not been described simultaneously by a single physical model, yet. In this research, we investigate whether recent avalanche models can describe the distributions for both the released energies and waiting times of flares in an active region. Flaring events are simulated using the modified Lu and Hamilton model (Lu and Hamilton (1991), ApJ, 380, 89) and also the optimized model (Farhang et al. (2018), ApJ, 859, 41). Applying a probability balance equation approach, we study the statistics of the simulated flaring events and investigate the origin of the observed power law in the flare frequency-size distribution. The results indicate that the power law originates in the distribution of transition rates (the distribution of the probabilities of transitions between different energies) rather than the distribution of the energy of the active region. It is also observed that the waiting-time distribution of simulated flaring events follows a \textit{q}-exponential function which approximates a simple Poisson distribution.
\end{abstract}

\section{Introduction}\label{sec:intro}
Solar flares are triggered by ohmic dissipation of electric current sheets (magnetic reconnection) in the solar atmosphere due to evolution of the Sun's complex magnetic field driven by emerging, canceling, twisting, and braiding of magnetic fields \citep{shibata2011, fletcher2011, Loureiro2016}. The excess magnetic energy stored in the current sheets channels into heating of coronal plasma, acceleration of charged particles, generation of shock waves, wave propagation, and mass ejections \citep{Aschwanden_2017}.

Solar flares are observed across the entire electromagnetic spectrum. Decades of observations indicate that solar flare energies span almost eight orders of magnitude, between
$ 10^{25} $
erg and
$ 10^{33} $
erg \citep{Crosby1993, shimizu1995, Aschw2000b, Aschwanden2014, maehara2015}. Regarding previous studies, the frequency-size distribution of flare energy
$ (E) $
follows a power law:
\begin{equation}
\label{eq1}
N(E) \varpropto E^{-\gamma},
\end{equation}
where
$ N(E) $
is the number of events per unit energy and per unit time, and
$ \gamma $
is the power-law index \citep{mike2001, Litvinenko2001, aschwbook2011, fletcher2011}.

The diversity of reports on observations of the solar flare waiting-time distribution (WTD) indicates the dependence of the WTD on factors like the time of the study, the choice of active region, and also the measure of the magnitude of flares \citep{boffeta1999, mike2001, Buchlin2005, Mike2008}. Observed WTDs of flaring events are found to follow either a simple Poisson or a time-dependent Poisson distribution. Study of solar flare waiting times over a long period also manifest a power-law behavior in the tail of the WTD. Introducing a physical model capable of describing both the solar flare energies and their waiting times is an important outstanding problem.

Inspired by the power-law behavior of the flare frequency-size distribution which originates in the scale-free and stochastic nature of these phenomena, the concept of self-organized criticality and the cellular automaton (CA) approach are widely applied in modeling solar flares. The underlying mechanism is described as an avalanche process \citep{LH1991, LH1993, zirker1993, robinson1994, isliker1998, boffeta1999, isliker2000, charbon2001, buchlin2003, hughes2003, barpi2007, morales2008b, stru2014, farhang2018}. \cite{LH1991} presented the first CA model for solar flares (hereafter the LH model) following the idea of the sandpile model \citep{bak1987}.

In the sandpile model, evolution of a discrete vector field is numerically studied subject to a constant-rate driving. If the driving mechanism causes an instability in the system the field is redistributed to locally relax the system. The LH model was successful in reproducing the power-law distribution for the size of simulated flaring events.

Other models have also been used to explain the power-law distribution. \cite{RV78} studied the frequency-size distributions of various transient sources (e.g., flaring stars) and perceived that the examined distributions follow a power-law behavior at high energies, but depart from the power law at low energies. \citeauthor{RV78} constructed a model for flaring events describing the power-law behavior and the observed departure. They assumed that the rate of energy storage (driving rate) is proportional to the energy of the system:
$ dE/dt = \alpha (E+E_{0}), $
where
$ E_{0} $
is the ground state energy. This implies an exponential growth of energy in the system. They also assumed flares occur as an uncorrelated process, i.e., the WTD of flaring events is a simple Poisson distribution with a constant flaring rate
$ (\lambda) $
:
\begin{equation}
\label{eq2}
P(\Delta t)=\lambda e^{-\lambda \Delta t},
\end{equation}
where
$ \Delta t $
is the time interval between two consecutive flaring events. With this model, the frequency-size distribution of events is:
\begin{equation}
\label{eqq2}
N(E) = \frac{N_{0}}{T}\frac{\lambda}{\alpha E_{0}}(1+E/E_{0})^{-\gamma},
\end{equation}
where
$ N _{0} $
and
$ T $
are the total number of flares and total time of observation, respectively. Equation (\ref{eqq2}) departs from a power law for small values of
$ E/E_{0}, $
and is a simple power law
$ \thicksim E^{-\gamma} $
for
$ E \gg E_{0}. $
\citeauthor{RV78} pointed out that a deviation from the exponential growth of energy results in a frequency-size distribution other than a power law.

\cite{Mike1998} presented a description of the energy of an active region in terms of a probability balance equation (master equation). In the model, flares represent transitions between energy states. They investigated whether the power-law flare size distribution originates in the distribution of the energy of the active region, or comes from a power law in the transition rates between energies of the active region. Their numerical analysis showed that an energy-independent driving rate can result in a power-law frequency-size distribution for flaring events. The power-law behavior was attributed to the transition rate function between the energy states. They argued that this is consistent with the avalanche model. They also noticed that the WTD of flaring events in the model is exponential, consistent with the avalanche model.

Here, we aim to study the energy balance in CA avalanche models, and investigate: 1. whether it is possible to calculate the transition rate in avalanche models, 2. whether the power-law behavior of flare frequency-size distribution originates in the transition rate, or reflects the distribution of the system energy in the avalanche models, and 3. whether avalanche models in fact have a simple Poisson distribution for the WTD. In Section \ref{sec:models}, we briefly review the CA avalanche models and the probability balance approach, and in Section \ref{sec:res} we discuss our numerical analysis and results. Finally, we present conclusions in Section \ref{sec:con}.

\section{Statistical Models}\label{sec:models}
\subsection{Cellular Automaton Avalanche Models}\label{sec:CA}
As the magnetic field lines and their surrounding plasma come out of the shearing layer of the tachocline and rise through the convection zone they become stretched and twisted. Some of these field lines can break through the surface of the Sun and create pairs of magnetic footpoints. The slow and continuous shuffling of magnetic footpoints due to photospheric motions increases magnetic stress in the coronal magnetic field. In case of an instability, magnetic reconnection occurs and the system locally relaxes by abrupt release of accumulated energy in the solar atmosphere. Each magnetic reconnection may trigger other instabilities in the system. The Sun's atmospheric magnetic field evolves quasi-statically towards a critical state that produce scale-free and nonlinear dissipation of energy \citep{Aschwanden2014}.

\cite{LH1991} established a CA model in order to simulate solar flaring events as a self-organized critical system. They designed a discrete 3D grid of nodes, using uniformly distributed random numbers, where the nodal values represent the average magnetic field
$ (\boldsymbol{B}) $
in each cell. The topological progression of the magnetic field is considered in the modeling as the system is subjected to a driving mechanism by adding one small random increment to a randomly selected node at each time step. Since the system is stipulated to remain stable, the stability of the whole system is checked after each driving step against a criterion:
\begin{equation}
\label{eq3}
\left| \triangle B_{i,j,k}\right| \hspace{1mm}\equiv \hspace{1mm} \left| B_{i,j,k} - \frac{1}{6}\sum_{l=1}^{6} B_{l}\right| > B_{c},
\end{equation}
where the sum runs over the six nearest neighbors
$ (B_{l}), $
and
$ B_{c} $
is a preset threshold. If the instability criterion exceeds the threshold at a node, the field redistributes as:
\begin{eqnarray}
\label{eq4}
B_{i,j,k}^{n+1}=&&B_{i,j,k}^{n}- \frac{6}{7} B_{c}, \vspace{13mm} \nonumber \\
B_{l}^{n+1}=&&B_{l}^{n}+ \frac{1}{7}B_{c}.
\end{eqnarray}

Several types of redistribution rules in various categories such as isotropic or anisotropic, deterministic or probabilistic, short-distance or long-distance interactions (involving different numbers of neighboring cells in the redistribution), and conservative or non-conservative have been subjected to a wide range of studies \citep{moore1962, Neumann66, Upper1997, weimar98, abigail2013, stru2014}. All redistribution rules in the existing CA models are ad hoc, and various models are credible.

Recently, the concept of the principle of minimum energy has been applied in modeling magnetic reconnections \citep{farhang2018, aschwanden_2018}. \cite{farhang2018} presented a CA model (hereafter the optimized model) in which the maximum possible amount of energy is released at each redistribution. They considered a 2D lattice of magnetic vector potential field
$ (\boldsymbol{A}), $
and instead of the local driving mechanism of the LH model, they applied a global driving procedure \citep{stru2014}:
\begin{equation}
\label{eq11}
A_{i,j}^{t+1}=(1+\epsilon)A_{i,j}^{t} \hspace{6mm} \forall (i,j),
\end{equation}
with the driving rate
$ \epsilon. $
The instability criterion is defined as:
\begin{equation}
\label{eq12}
\left| \triangle A_{i,j} \right| \equiv \left| A_{i,j} - \frac{1}{4}\sum_{k=1}^{4} A_{k}\right| > A_{c},
\end{equation}
where
$ A_{c} $
is the instability threshold generated from a Gaussian distribution, and the sum runs over the four nearest neighbors. Among all unstable sites, only nodes that could deplete the maximum possible amount of energy are redistributed as following:
\begin{eqnarray}
\label{eq13}
A_{i,j}^{n+1}=&&A_{i,j}^{n}- \frac{4}{5} A_{c}, \vspace{13mm} \nonumber \\
A_{i,j-1}^{n+1}=&&A_{i,j-1}^{n}+ \frac{4}{5}\frac{r_{1}}{x+a}A_{c}, \vspace{16mm} \nonumber \\
A_{i+1,j}^{n+1}=&&A_{i+1,j}^{n}+ \frac{4}{5}\frac{r_{2}}{x+a}A_{c}, \vspace{16mm} \nonumber \\
A_{i,j+1}^{n+1}=&&A_{i,j+1}^{n}+ \frac{4}{5} \frac{r_{3}}{x+a}A_{c}, \vspace{16mm} \nonumber \\
A_{i-1,j}^{n+1}=&&A_{i-1,j}^{n}+ \frac{4}{5} \frac{x}{x+a}A_{c}.
\end{eqnarray}
Here
$ r_{1}, r_{2}, r_{3}  $
are uniformly distributed random numbers,
$ a=r_{1}+r_{2}+r_{3}, $
and
$ x $
is a free parameter that is determined using the principle of minimum energy (see Appendix \ref{sec:appendix1}).

In the conservative anisotropic redistribution rules of Equation (\ref{eq13}), a fraction of the field is subtracted from the central node and is redistributed to the neighboring cells in order to maximize the released energy.

Another approach to achieving maximum energy release is to introduce an isotropic set of redistribution rules:
\begin{eqnarray}
\label{eq14}
A_{i,j}^{n+1}=&&A_{i,j}^{n}- \frac{4}{5} A_{c} x, \nonumber \\
A_{i\pm 1,j\pm 1}^{n+1}=&&A_{i\pm 1,j\pm 1}^{n}+ \frac{1}{5} A_{c} x,
\end{eqnarray}
which is conservative as well. Equation (\ref{eq14}) is the same as the 2D LH model if
$ x=\frac{\mid\triangle A_{i,j}\mid}{A_{c}}=1, $ but for
$ x>1 $
the amount of decrease in
$ A^{2} $
due to the occurrence of each redistribution, equivalent to the energy loss in the system equal to
$ 4/5 z_c^{2} \left( 2\frac{\mid\triangle A_{i,j}\mid}{A_{c}}x - x^{2} \right), $
is always greater than the LH model (see Appendix \ref{sec:appendix2}).

\subsection{Probability Balance Approach}\label{sec:ME}
The steady state of a system comprised of a continuous range of possible energy states can be described by a probability balance equation \citep{vankamp1992, Mike1998, Mike2008}:
\begin{equation}
\label{eq5}
\frac{d}{dE}(\dot{E}P)+P \int_{0}^{E} \alpha(E,E^{'})dE^{'} - \int_{E}^{\infty} P(E^{'})\alpha(E,E^{'})dE^{'} =0,
\end{equation}
where
$ P(E)dE $
is the probability of the system having energy in the interval
$ (E,E+dE), \dot{E} $
is the driving rate, and
$ \alpha(E,E^{'})dE^{'} $
is the probability per unit time of the system with energy
$ E $
making a transition to a lower energy level in the interval
$ (E^{'},E^{'}+dE^{'}). $
The first term of the probability balance equation describes the increase of energy due to driving, the second term describes the change in energy due to the system falling from energy
$ E $
to a lower energy, and the third term describes the system falling from a higher energy to energy
$ E. $

The frequency-size distribution of transitions is:
\begin{equation}
\label{eq6}
N(E)=\int_{E}^{\infty} P(E^{'})\alpha(E^{'},E^{'}- E)dE^{'},
\end{equation}
and the total occurrence rate of transitions is:
\begin{equation}
\label{eq7}
\lambda(E) =\int_{0}^{E} \alpha(E,E^{'})dE^{'}.
\end{equation}

\cite{Mike2008} applied this concept to solar flares occurring in an active region. Equation (\ref{eq7}) indicates that in general the occurrence (flaring) rate depends on the energy
$ (E) $
of the system. Therefore, one might think of solar flares as correlated processes whereas on the contrary, \cite{mike2002wt} assumed that the WTD of solar flares is a Poisson distribution.

\cite{Mike2008} investigated a numerical steady state solution to Equation (\ref{eq7}) for solar flares, assuming a constant driving rate
$ \dot{E}. $
\citeauthor{Mike2008} also assumed a power-law like transition rate:
\begin{equation}
\label{eq8}
\alpha(E,E^{'})=\alpha_{0} E^{\delta} (E-E')^{-\gamma} \theta(E-E'-E_{0}),
\end{equation}
where
$ \theta $
is the step function, and the size of a flaring event (the released energy) is
$ E-E^{'}\geq E_{0}. $
Equation (\ref{eq8}) provides the potential for dependence of the transition rate on the excess energy of the system through the factor
$ E^{\delta}, $
if
$ \delta \neq 0. $
Substituting the transition rate of Equation (\ref{eq8}) in Equations (\ref{eq6}) and (\ref{eq7}) gives:
\begin{eqnarray}
\label{eq9}
N(E)=\alpha_{0} E^{-\gamma} \int_{E}^{\infty} (E^{'})^{\delta} P(E^{'}) dE^{'},
\end{eqnarray}
and
\begin{eqnarray}
\label{eq10}
\lambda(E) =\frac{\alpha_{0}}{1-\gamma} E^{\delta} (E^{1-\gamma}-E_{0}^{1-\gamma}).
\end{eqnarray}

Equation (\ref{eq9}) indicates that the flare frequency-size distribution in the model is a power-law function with index
$ \gamma, $
times the function
$ \int_{E}^{\infty} (E^{'})^{\delta} P(E^{'}) dE^{'}. $

Equation (\ref{eq10}) implies that the total flaring rate
$ \lambda(E) $
is a fluctuation of the energy of the system. This indicates that WTD will not be a simple exponential (which requires
$ \lambda = \lambda_{0}, $
a constant). However, the WTD will be approximately exponential if the system energy
$ E $
is much greater than the energy of flares. Specifically, for
$ E \gg E_{0} $
and
$ \gamma > 1 $
Equation (\ref{eq10}) implies:
\begin{eqnarray}
\label{eqqq10}
\lambda(E) \approx \frac{\alpha_{0}}{\gamma-1} E^{\delta} E_{0}^{1-\gamma},
\end{eqnarray}
and if
$ E $
greatly exceeds the flare energies then
$ E \approx \bar{E}, $
the average system energy, and
\begin{eqnarray}
\label{eqqqq10}
\lambda(E) \approx \frac{\alpha_{0}}{\gamma-1} \bar{E}^{\delta} E_{0}^{1-\gamma}=\textrm{const}.
\end{eqnarray}
With the same assumptions, we expect
$ P(E) $
to be a sharply peaked distribution around the average system energy, i.e.,
\begin{eqnarray}
\label{eqqqqq10}
P(E) \approx \delta (E - \bar{E}),
\end{eqnarray}
and then Equation (\ref{eq9}) becomes:
\begin{equation}
N(E) \approx \left\{
\begin{array}{rl}
\alpha_{0}\bar{E}^{\delta}E^{-\gamma} & \text{for } E \leq \bar{E},\\
 0~~~~~ & \text{for } E > \bar{E},
\end{array} \right.
\end{equation}
i.e. the frequency-energy distribution for flares is a simple power law.
\section{Analysis \& Results}\label{sec:res}
In this section we present numerical analysis to determine the frequency-size distribution and also the WTD of flaring events in recent avalanche models. The dimensionality of the LH model is reduced by considering a 2D lattice of magnetic vector potential field
$(\boldsymbol{A})$
in the cross section of a flaring magnetic flux tube \citep{stru2014}. The lattice is driven by adding a uniformly distributed random number to a random node. The driving
$ (\mathrm{\epsilon_{LH}}) $
is slow in the sense that
$ \mathrm{\epsilon_{LH}}/\langle A \rangle < 10^{-4}. $
The lattice energy is considered proportional to
$ A^{2}. $

After each driving step the stability of the system is checked against the instability criterion:
\begin{equation}
\label{eq15}
\left| \triangle A_{i,j}\right| \hspace{1mm} \equiv \hspace{1mm} \left| A_{i,j} - \frac{1}{4}\sum_{l=1}^{4} A_{l}\right| > A_{c}.
\end{equation}
The instability threshold
$ A_{c} $
is adopted from a Gaussian distribution with an average of
$ \bar{A_{c}}=1, $
and full width at half-maximum
$ \sigma=0.01. $
In case of an instability, a redistribution takes place and the system locally relaxes. Each redistribution may lead to other instabilities in the system. A succession of redistributions is called an avalanche (flare).
%A succession of redistributions from an inceptive instability somewhere to an ultimate stability everywhere is called an avalanche (flare).

We present simulations for the optimized model for both types of redistribution rules described in Section \ref{sec:models}: the anisotropic set of Equations (\ref{eq13}), and the isotropic set of Equations (\ref{eq14}). In both cases, a 2D lattice of magnetic vector potential field
$ (\boldsymbol{A}) $
is built up using uniformly distributed random numbers. For the anisotropic case, the lattice is subject to the global driving mechanism (Equation (\ref{eq11})). The magnetic energy of the lattice is calculated using the following expression:
\begin{eqnarray}
\label{eq16}
E=\frac{1}{2} \sum \boldsymbol{A}\cdot\boldsymbol{J}.
\end{eqnarray}

The isotropic case is developed applying some variations from the original optimized model. Firstly, the system is subject to a local driving mechanism instead of global driving. Moreover, the lattice energy is considered proportional to
$ A^{2}, $
in common with the reduced LH model. In the reduced LH model the stationary state is achieved after nearly
$ 20000 $
driving steps, whereas in the optimized model it takes several million iterations for the system to reach a stationarity state.

\begin{figure}
\centering
\includegraphics[scale=0.45]{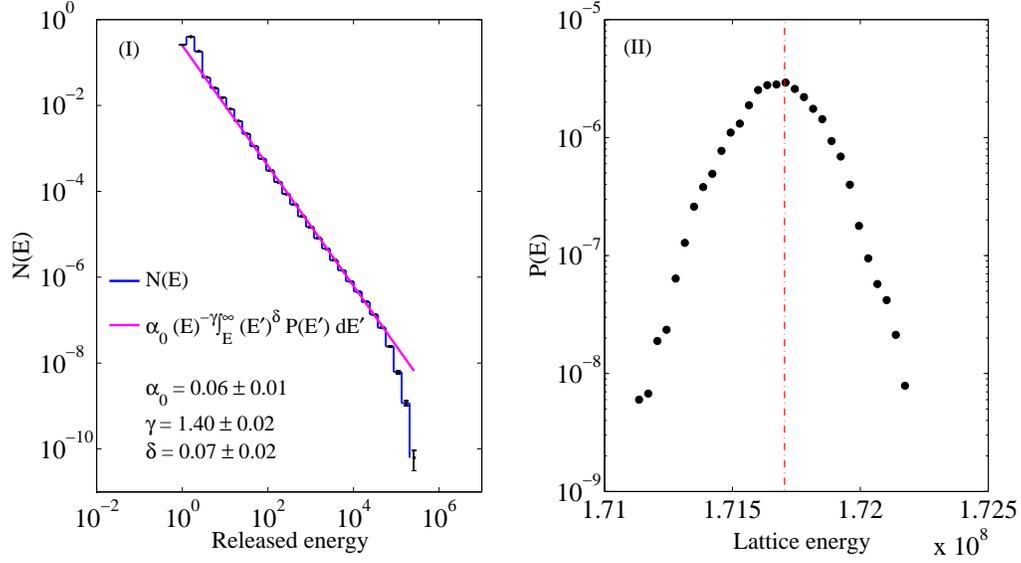}
\caption{The statistics of the simulated flaring events for the reduced LH model. (I) The frequency-size distribution of flaring events together with the result of fitting the Equation (\ref{eq9}) on the data. (II) The probability distribution of lattice energies. The location of the peak of $ P(E) $ is shown with the red line.}
\label{fig1}
\end{figure}

\begin{figure}
\centering
\includegraphics[scale=0.45]{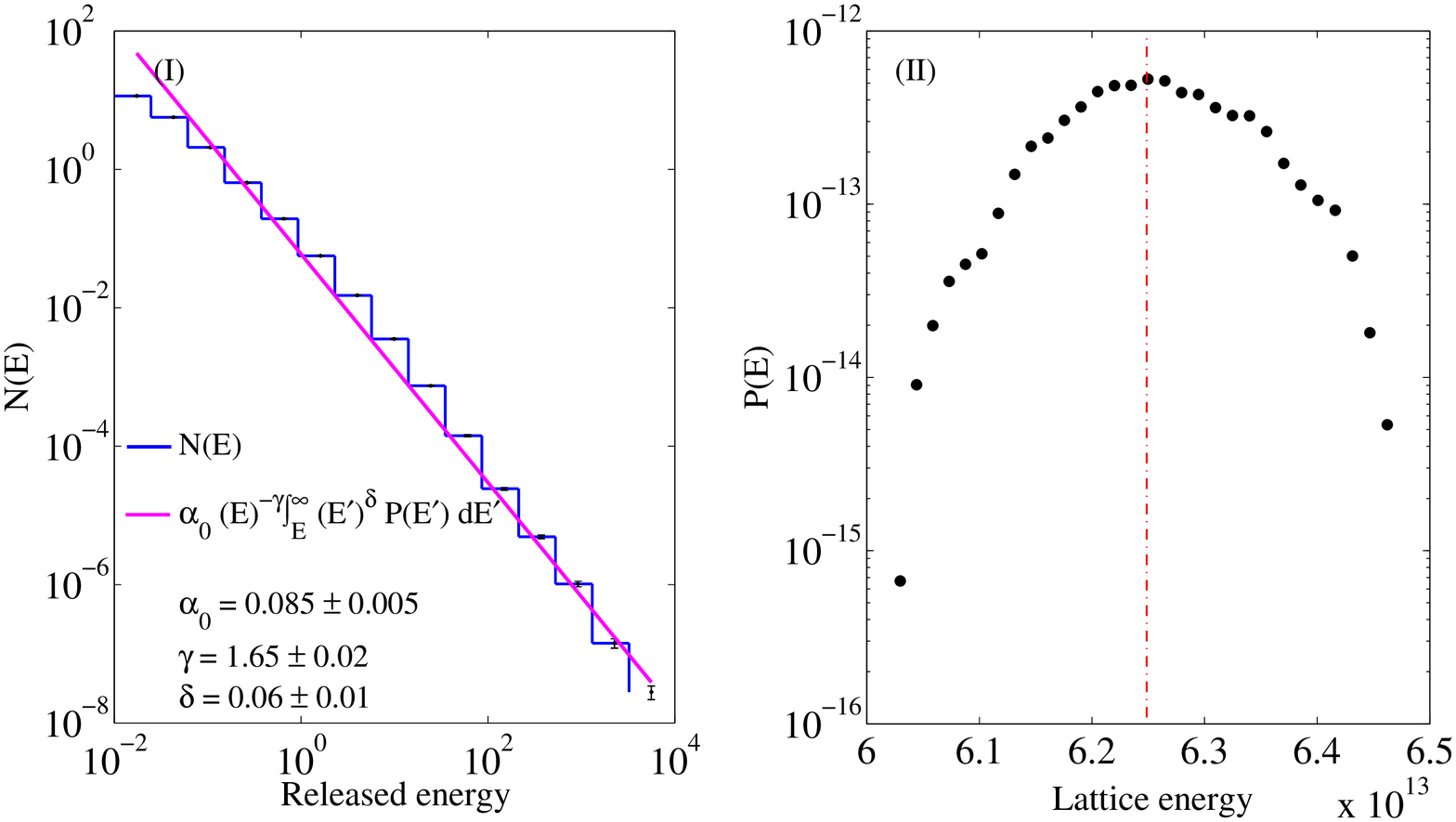}
\caption{The statistics of the simulated flaring events for the anisotropic optimized model. (I) The frequency-size distribution of flaring events together with the result of fitting the Equation (\ref{eq9}) on the data. (II) The probability distribution of lattice energies. The location of the peak of $ P(E) $ is shown with the red line.}
\label{fig2}
\end{figure}

\begin{figure}
\centering
\includegraphics[scale=0.45]{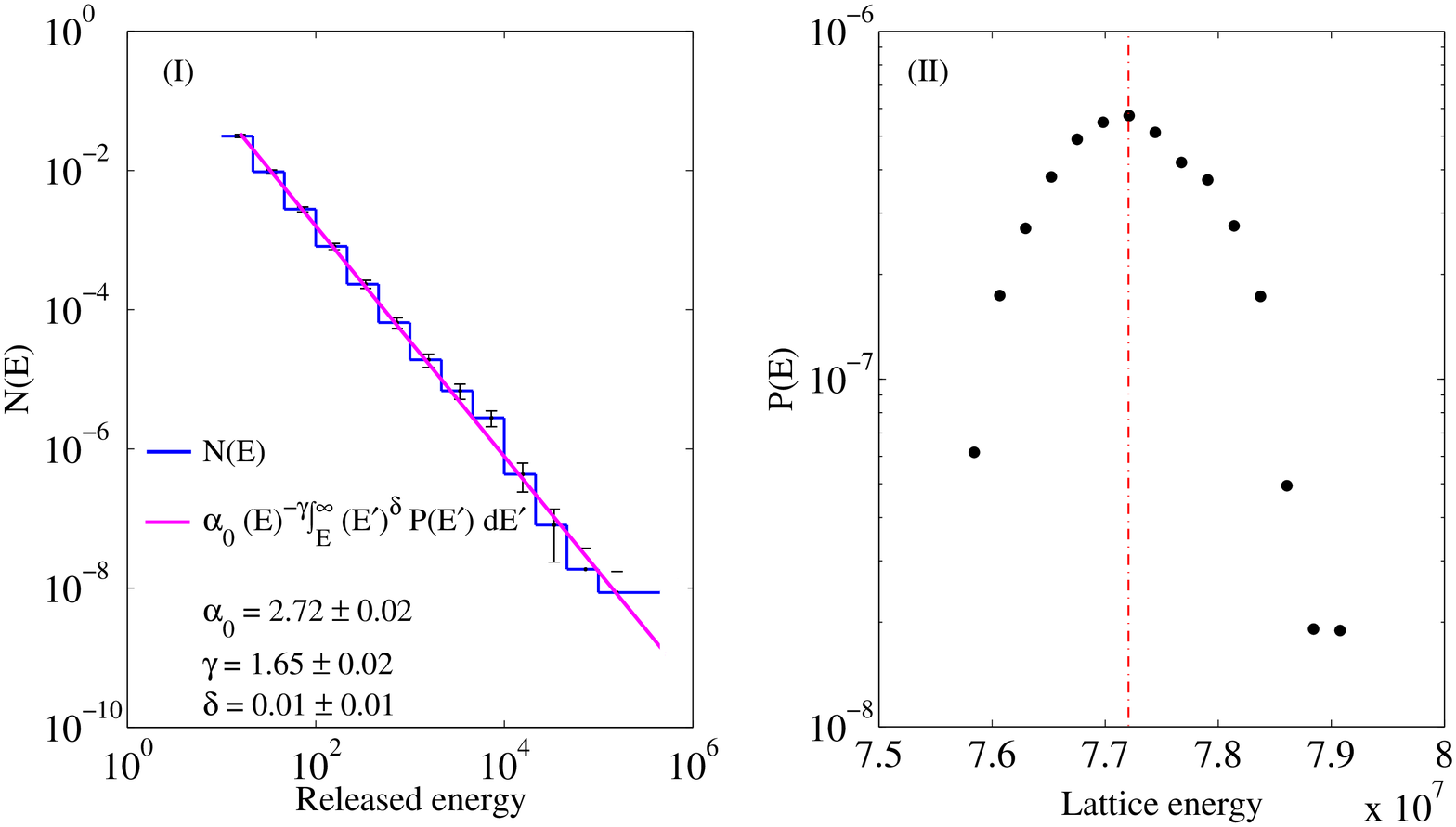}
\caption{The statistics of the simulated flaring events for the isotropic optimized model. (I) The frequency-size distribution of flaring events together with the result of fitting the Equation (\ref{eq9}) on the data. (II) The probability distribution of lattice energies. The location of the peak of $ P(E) $ is shown with the red line.}
\label{fig3}
\end{figure}

The frequency-size distributions of the simulated flares together with the probability distribution functions (PDF) of lattice energies for the reduced LH model and the optimized model are shown in Figures \ref{fig1}-\ref{fig3}. The maximum released energy (in the left panels) is much smaller than
 $ \bar{E} $
(the location of peak in the right panels) over several million iterations of simulation. This is due to the finite time of simulation. Specifically, the expected number of events with size greater than
$ E $
during the total simulation time
$ T $
is:
\begin{eqnarray}
\label{eqq16}
n(>E) = T \int_{E}^{\infty} N(E') dE',
\end{eqnarray}
where
$ N(E) $
is the flare frequency distribution. The value of
$ E $
at which
$ n(>E) = 1, $
represents the energy rollover in the system as a result of the finite time of the simulation. The rollovers observed in Figures \ref{fig1}-\ref{fig3} correspond to
$ n = 1. $

In order to calculate the transition rate in the numerical models, we assume that transition rate depends on the lattice energies and also released energies according to Equation (\ref{eq8}). Therefore, fitting the right hand side of Equation (\ref{eq9}) to the frequency-size distributions of individual data sets provides the transition rate. The goodness of fit is evaluated using the
$ \chi^{2} $
test. Hence, we apply the genetic algorithm \citep{canto2009, kramer2017, farhang2018} to minimize the
$ \chi^{2} $
function with respect to parameters
$ \alpha_{0}, \gamma, $
and
$ \delta $
to obtain their values consistent with our distributions. We also use the Monte Carlo integration technique to calculate the factor
$ \int_{E}^{\infty} (E^{'})^{\delta} P(E^{'}) dE^{'} $
over the lattice energies.

The quasi-normal shape of the PDFs after the system reaches to the stationary state (right panels in Figures \ref{fig1}-\ref{fig3}) is consistent with Equation (\ref{eqqqqq10}), and clearly reveals that the distribution of energy
$ P(E) $
is not the origin of the power-law frequency distribution of event energies. This result confirms that the power-law behavior must originate in the transition rates.

\begin{figure}
\centering
\includegraphics[scale=0.5]{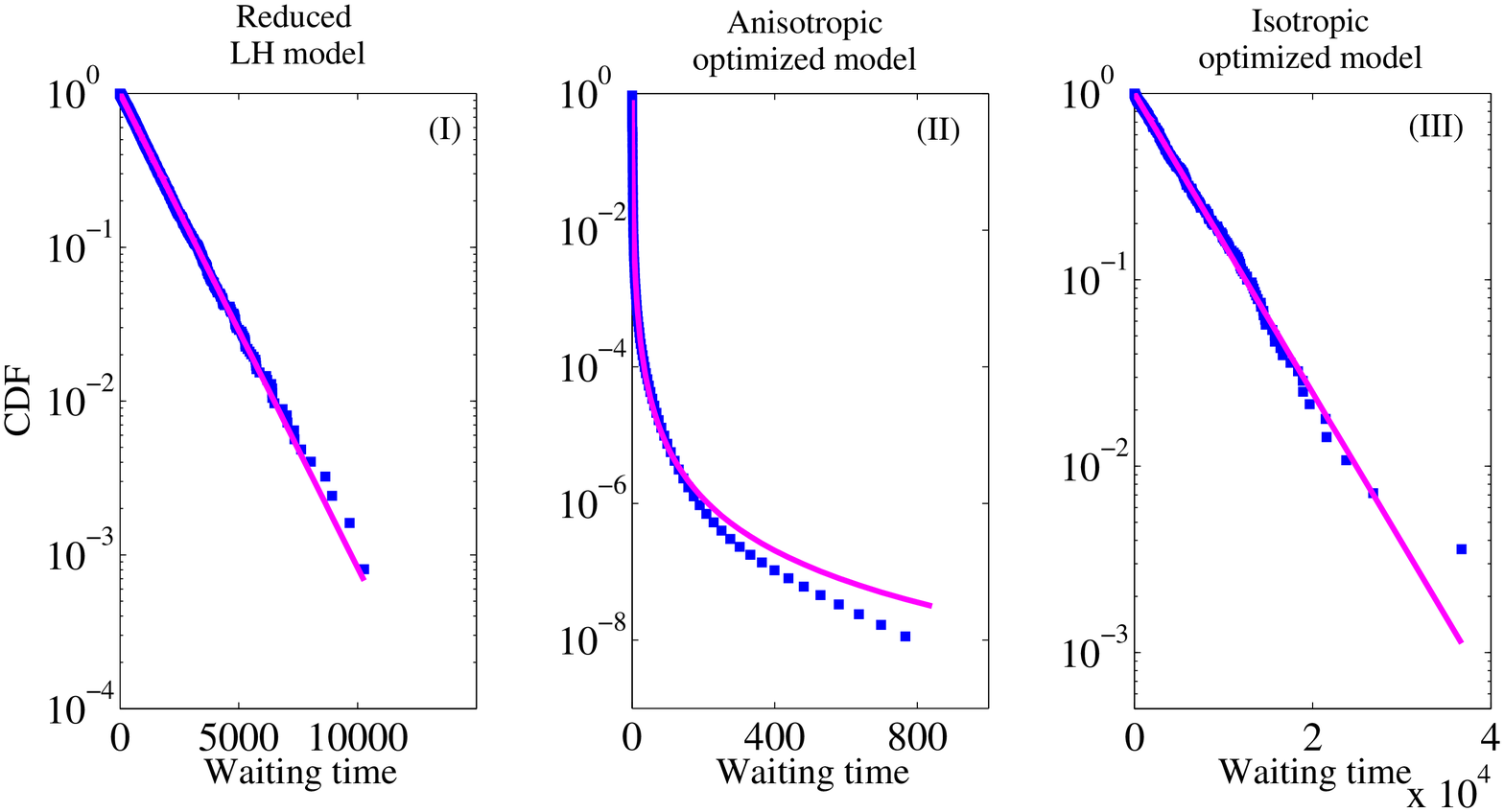}
\caption{The cumulative waiting-time distributions for the reduced LH model (I), the anisotropic optimized model (II), and the isotropic optimized model (III), respectively. The red line in each panel represents the cumulative \textit{q}-exponential distribution of Equation (\ref{eqqexp}). The obtained values for $ q $ are
$ 1.00001 \pm 0.00001, 1.28 \pm 0.02, $
and
$ 1.00010 \pm 0.00001, $
respectively.}
\label{fig5}
\end{figure}

Observations suggest that solar flare WTD in individual active regions is basically exponential \citep[e.g.,][]{mike2002wt}. Also, the study of solar flare waiting times for the whole Sun, i.e., over many active regions indicates a power-law behavior in the tail of the WTD \citep{McTiernan2010}. \cite{gheibi2017} discussed the long-term dependency detected in the system of flares (flare time-series), which is a key characteristic of complex systems. The WTD of such systems could be well-described by the \textit{q}-exponential probability distribution function \citep{Tsallis2004, Yalcin2016}:
\begin{eqnarray}
\label{eq17}
%e^{\lambda \Delta t} = Q_{0} (1 + \lambda (q-1) \Delta t)^{-\gamma},
P(\Delta t)=(2-q)\lambda e^{-\lambda \Delta t} \approx (2-q)\lambda \bigg( 1 - (1-q)(\lambda \Delta t) \bigg)^{\frac{1}{1-q}}.
\end{eqnarray}
Equation (\ref{eq17}) is valid in the range
$ 0\leq \Delta t < \infty $
for
$ q \geq 1, $
otherwise
$ (q < 1) $
it is only justifiable in the range
$ 0\leq \Delta t < \frac{1}{\lambda (1-q)} $
\citep{qex2009}. The simple exponential distribution (ordinary exponential distribution
$ \lambda e^{-\lambda \Delta t} $
) is achieved as
$ q $
approaches unity. The corresponding cumulative distribution function (CDF) of Equation (\ref{eq17}) is:
\begin{eqnarray}
\label{eqqexp}
C(\Delta t)= \bigg( 1- (1-q)(\lambda \Delta t) \bigg)^{1+\frac{1}{1-q}}.
\end{eqnarray}
The cumulative waiting-time distributions are fitted by Equation (\ref{eqqexp}) and the results are shown in Figure \ref{fig5}. The obtained values for
$ q $
are
$ 1.00001 \pm 0.00001, 1.28 \pm 0.02, $
and
$ 1.00010 \pm 0.00001 $
for the reduced LH, anisotropic, and isotropic optimized models, respectively. Expectedly, the WTD for the reduced LH model is perfectly matched with the simple Poisson distribution. The results also show that the WTD for the isotropic optimized model is consistent with the simple Poisson distribution, but the anisotropic model departs from the simple Poisson form.

\section{Conclusion}\label{sec:con}
In this study, the complex evolution of the coronal magnetic field is considered as a probabilistic combination of energy states. Namely, it is assumed that the coronal magnetic field consists of a continuous distribution of energy states, and transitions can occur between the states at different rates (due to e.g., flaring events or a driving mechanism). The evolution of such a system is described by the probability balance equation. In this paper we have investigated the energy balance in CA avalanche models, using the probability balance equation approach.

To this end, a 2D lattice of magnetic field (magnetic vector potential field) was constructed representing a cross section of a typical flaring flux tube within an active region. Uniformly distributed random numbers were assigned to the nodal values of the lattice, as a small fraction of the average of the magnetic vector potential field within each cell. Both the local and the global driving mechanisms were investigated. The behavior of the flaring event energies and their waiting times are analysed for the reduced LH model and also both types of optimized model.

Two important questions are addressed here. The first concerns the flare energies. Does the power-law behavior in the frequency-size distribution of flaring events originate in the transition rates between energy states, or it is due to the distribution of system energy? To answer this question, the transition rate was calculated for each avalanche model. The probability distribution functions for the lattice energies are found to exhibit a quasi-normal behavior. The results demonstrate that a power-law distribution for the transition rates can explain the observed flare frequency-energy distribution. This indicates that the power-law behavior originates in the distribution of transition rates rather than the distribution of energy of the active region.

The second question concerns the WTD of flaring events. \cite{mike2002wt} stated that the WTD of the solar flares could be described by a Poisson distribution. The question is whether the avalanche models provide the same result. We investigated the WTD of the simulated flaring events, applying a \textit{q}-exponential distribution. The \textit{q}-exponential distribution exhibits the same behavior as the simple Poisson distribution for
$ q=1, $
which is obtained for the reduced LH model and also the isotropic optimized model. In case of the anisotropic optimized model, the obtained value for
$ q $
shows some deviation from unity. These results indicate that the WTD of flaring events in the avalanche models can be a simple Poisson distribution, or it may be more complex.

In the next logical step, our interest is to examine the mentioned models for the observational total magnetic energy of active regions and flare energies using recent data recorded by e.g., the Atmospheric Imaging Assembly (AIA) and Helioseismic and Magnetic Imager (HMI) on board the Solar Dynamics Observatory (SDO). This could help us determine a more physical avalanche model.
\newpage
\appendix
\section{Optimization of The Released Energy for the anisotropic model}\label{sec:appendix1}
In order to determine the value of
$ x $
in Equation (\ref{eq13}), which maximizes the amount of released energy, the first derivative of the energy difference between two consecutive driving steps,
$ E^{n+1}-E^{n}, $
is calculated. Considering the energy of the system as
$ E=\sum \frac{1}{2} \boldsymbol{A}\cdot\boldsymbol{J}, $
and also assuming that the released energy during a redistribution depends on the nodal values of all 13 neighboring
cells \cite[Figure 1 therein]{farhang2018}, then
$ x $
is determined by solving:
\begin{eqnarray}
\label{eqap4}
\frac{\partial (E^{n+1}-E^{n})}{\partial x}=&& \frac{\partial J_{i,j-2}^{n+1}}{\partial x} A_{i,j-2}^{n+1} + \frac{\partial J_{i-1,j-1}^{n+1}}{\partial x} A_{i-1,j-1}^{n+1}+\frac{\partial J_{i,j-1}^{n+1}}{\partial x} A_{i,j-1}^{n+1} \nonumber\\
+&& \frac{\partial J_{i+1,j-1}^{n+1}}{\partial x} A_{i+1,j-1}^{n+1}+\frac{\partial J_{i-2,j}^{n+1}}{\partial x} A_{i-2,j}^{n+1} +\frac{\partial J_{i-1,j+1}^{n+1}}{\partial x} A_{i-1,j+1}^{n+1} \nonumber\\
+&&\frac{\partial J_{i+2,j}^{n+1}}{\partial x} A_{i+2,j}^{n+1}+\frac{\partial J_{i+1,j+1}^{n+1}}{\partial x} A_{i+1,j+1}^{n+1}+\frac{\partial J_{i,j+2}^{n+1}}{\partial x} A_{i,j+2}^{n+1}\nonumber\\
+&&\frac{\partial J_{i,j+1}^{n+1}}{\partial x} A_{i,j+1}^{n+1}+\frac{\partial J_{i-1,j}^{n+1}}{\partial x} A_{i-1,j}^{n+1} +\frac{\partial J_{i+1,j}^{n+1}}{\partial x} A_{i+1,j}^{n+1}\nonumber\\
+&&J_{i,j-1}^{n+1} \frac{\partial A_{i,j-1}^{n+1}}{\partial x}+ J_{i,j+1}^{n+1} \frac{\partial A_{i,j+1}^{n+1}}{\partial x}+ J_{i+1,j}^{n+1} \frac{\partial A_{i+1,j}^{n+1}}{\partial x}\nonumber\\
+&& J_{i-1,j}^{n+1} \frac{\partial A_{i-1,j}^{n+1}}{\partial x} +\frac{\partial J_{i,j}^{n+1}}{\partial x} A_{i,j}^{n+1}=0,
\end{eqnarray}
where the nondimensional electric current density is:
\begin{eqnarray}
\label{eqap1}
\boldsymbol{J}=\boldsymbol{\nabla}\times \boldsymbol{B}= \boldsymbol{\nabla}\times (\boldsymbol{\nabla}\times\boldsymbol{A}).
\end{eqnarray}
The magnetic vector potential is defined as
$ \boldsymbol{A}=A(x,y)\hat{z}, $
hence
$ \boldsymbol{\nabla}\cdot\boldsymbol{A}=0, $
and we have:
\begin{eqnarray}
\label{eqap2}
\boldsymbol{J}=-\left( \frac{\partial^{2}}{\partial x^{2}} + \frac{\partial^{2}}{\partial y^{2}} \right) \boldsymbol{A}.
\end{eqnarray}
Applying a centered difference approximation, the vertical component of the nondimensional electric current density is:
\begin{eqnarray}
\label{eqap3}
J_{z}\bigg|_{i,j}= 4A_{i,j}-A_{i+1,j}-A_{i-1,j}-A_{i,j+1}-A_{i,j-1}.
\end{eqnarray}
After some calculations:
\begin{eqnarray}
\label{eqap6}
\frac{\partial (E^{n+1}-E^{n})}{\partial x} = \frac{4}{10} \frac{A_{c}}{(x+a)^{2}} \bigg( &&r_{1} \big( -2J_{i,j-1}^{n}+3A_{i,j-1}^{n} \big) +r_{2} \big( -2J_{i+1,j}^{n}+3A_{i+1,j}^{n} \big)\nonumber \\
 +&& r_{3} \big( -2J_{i,j+1}^{n}+3A_{i,j+1}^{n} \big) + a \big( 2J_{i-1,j}^{n}-3A_{i-1,j}^{n}  \big) \nonumber \\
 -&& \frac{32}{5}\frac{Z_c}{x+a}\big( r_{1}^{2}+r_{2}^{2}+r_{3}^{2}-ax \big) \bigg)=0,
\end{eqnarray}
which leads to:
\begin{eqnarray}
\label{eqap7}
x=\frac{A_{c}(r_{1}^{2}+r_{2}^{2}+r_{3}^{2}-ax)-\frac{5}{32}a \Theta}{\frac{5}{32} \Theta + a A_{c}},\nonumber
\end{eqnarray}
and
\begin{eqnarray}
\Theta=&& r_{1} \big( -2J_{i,j-1}^{n}+3A_{i,j-1}^{n} \big) +
r_{2} \big( -2J_{i+1,j}^{n}+3A_{i+1,j}^{n} \big)\nonumber \\
+&&r_{3} \big( -2J_{i,j+1}^{n}+3A_{i,j+1}^{n} \big)
+a \big( 2J_{i-1,j}^{n}-3A_{i-1,j}^{n}  \big).
\end{eqnarray}
\section{Optimization of The Released Energy for the isotropic model}\label{sec:appendix2}
In the isotropic optimized model the lattice energy is considered proportional to
$ \sum A^{2}. $
Therefore, the released energy depends on the nodal values of 5 cells involved in the redistribution. Following the same procedure in Appendix \ref{sec:appendix1} the value of
$ x $
in Equation (\ref{eq14}) satisfying the maximum energy release constraint is determined by solving:
\begin{eqnarray}
\label{eqap8}
\frac{\partial (E^{n+1}-E^{n})}{\partial x}=\frac{\partial}{\partial x} \bigg( \frac{4}{5}x^{2}A_{c}^{2}-\frac{2}{5}x A_{c} J_{i,j}^{n} \bigg)= 0,
\end{eqnarray}
where
$ J_{i,j}^{n}=4 \triangle A_{i,j}. $
Equation (\ref{eqap8}) gives:
\begin{eqnarray}
\label{eqap9}
x=\frac{\mid \triangle A_{i,j} \mid}{A_{c}}.
\end{eqnarray}

\clearpage

\clearpage
\bibliography{ref_rep.bib}

\end{document}